# Pure-state single-photon wave-packet generation by parametric down conversion in a distributed microcavity

M. G. Raymer[*] and Jaewoo Noh[†]

*Oregon Center for Optics and Department of Physics, University of Oregon, Eugene, Oregon 97403*

K. Banaszek, I.A. Walmsley

*Clarendon Laboratory, University of Oxford, Parks Road, Oxford OX1 3PU, United Kingdom*

**Abstract**

We propose an optical parametric down conversion (PDC) scheme that does not suffer a trade-off between the state-purity of single-photon wave-packets and the rate of packet production. This is accomplished by modifying the PDC process by using a microcavity to engineer the density of states of the optical field at the PDC frequencies. The high-finesse cavity mode occupies a spectral interval much narrower than the bandwidth of the pulsed pump laser field, suppressing the spectral correlation, or entanglement, between signal and idler photons. Spectral filtering of the field occurs prior to photon creation rather than afterward as in most other schemes. Operator-Maxwell equations are solved to find the Schmidt-mode decomposition of the two-photon states produced. Greater than 99% pure-state packet production is predicted to be achievable.

---

[*] Electronic address: raymer@uoregon.edu

[†] Permanent address: Physics Department, Inha University, Inchon 402-751, Korea

**I - Introduction**

Quantum information processing and quantum communication can, in principle, be implemented by the use of linear optics and single-photon wave packets [1]. Such an implementation requires the use of many identical, synchronized, pure-state, single-photon wave packets. In a one-dimensional description, a pure-state, single-photon wave-packet state $|1\rangle_\psi$ is a superposition of monochromatic, single-photon states $|\omega\rangle = \hat{a}^\dagger(\omega)|vac\rangle$, weighted by an amplitude $\psi(\omega)$, [2]

$$|1\rangle_\psi = \int d\omega\, \psi(\omega)\, |\omega\rangle \quad . \tag{1}$$

The operator $\hat{a}^\dagger(\omega)$ creates a photon at angular frequency $\omega$. Such pure-state packets allow high-visibility interference of quantum amplitudes when two such photons come together at a beam splitter. The Bose symmetry of the photon states (or equivalently the commutator algebra of the field annihilation and creation operators) leads to unique interference effects, which allow the conditional operation of quantum logic gates. On the other hand, if the photons are created in mixed wave-packet states, described by a density operator,

$$\hat{\rho} = \sum_\psi P(\psi)\, |1\rangle_\psi{}_\psi\langle 1| \quad , \tag{2}$$

then the interference of two single-photon wave packets has low visibility, preventing scaling up of the system to many photonic qubits. Therefore, it is important to develop techniques for creating pure-state, single-photon wave-packets of the form Eq.(1).

Two primary techniques are being pursued for creating pure-state, single-photon wave-packets – single-atom (or quantum dot) [3], and spontaneous parametric down conversion (PDC) [4, 5]. In PDC – the topic of the present study – a "blue" pump laser field passes through a second-order nonlinear optical crystal (or a third-order optical fiber [6]), and, with small probability, a blue photon is converted into a pair of "red" photons. If the blue pump laser field (with frequency $\omega_p$) is continuous-wave and idealized as monochromatic, then the two red photons are perfectly correlated in frequency, with frequencies $\omega$ ("signal") and $\omega_p - \omega$ ("idler") [7]. In this case, the state of the red, down-converted field is [8]

$$|\Psi\rangle = |vac\rangle + \int d\omega \, C(\omega)|\omega\rangle_S |\omega_p - \omega\rangle_I \qquad , \qquad (3)$$

where $C(\omega)$ is the quantum amplitude for creation of the pair. Multi-pair creation can be ignored if the pump field is weak. If an idler photon with frequency precisely equal to $\omega_0$ is detected, then the state of the signal field created by this conditional process is

$$|\Psi\rangle_S = |\omega_p - \omega_0\rangle_S \quad , \qquad (4)$$

that is, an idealized monochromatic field at frequency $\omega_p - \omega_0$.

In order to operate a clocked, many-gate information-processing system, one needs temporally localized and synchronized photon packets. It is necessary to use a *pulsed* blue pump field to drive each down-conversion crystal, in order to synchronize the signal photons. This, however, leads to a nonzero pump spectral width, and destroys the perfect frequency correlation between signal and idler photons that exists in the case of a monochromatic pump (Eq.(3)).

For concreteness, consider the case of type-II PDC, where the signal and idler fields are spatially separable by their state of polarization (ordinary or extraordinary). See Fig.1. (In the case of type-I PDC one uses the direction of propagation to separate the pair.)

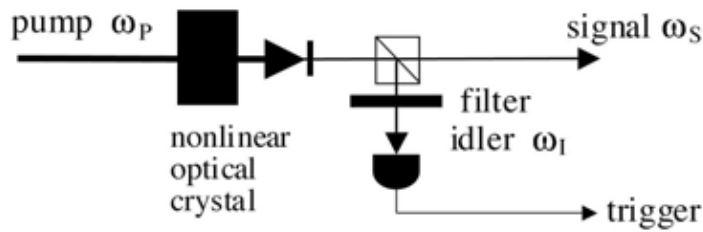

Fig.1 Conventional scheme, in which a laser pulse pumps a type-II optical parametric crystal. Signal and idler fields are produced and spatially separable by their state of polarization using a polarizing beam splitter. If an idler photon is detected after a narrow spectral filter, then the corresponding signal photon is in a nearly pure state of known frequency.

A narrow-band spectral filter, with line width $\Gamma$ and center frequency $\omega_0$, is placed into the idler beam, in front of a photon-counting detector, which serves as a trigger. When the idler trigger fires, one knows that a single photon is present in the signal beam (assuming that either the detector resolves photon number or that the probability of two pairs is vanishingly small). A strong test of the purity of single-photon packets would be to create two such packets in independent setups, and to interfere the signal photons from each source at a beam splitter, conditioned on recording idler trigger counts in both setups.

When the pump is pulsed, and therefore non-monochromatic, the state of the PDC field is

$$\left|\Psi\right\rangle = \left|vac\right\rangle + \int d\omega \int d\omega' C(\omega,\omega') \left|\omega\right\rangle_S \left|\omega'\right\rangle_I \quad , \tag{5}$$

where $C(\omega,\omega')$ is the quantum amplitude for creation of a photon pair with frequencies $\omega, \omega'$. Its form is determined by the pump spectrum and the phase-matching constraints for PDC [9]. As in [5], if one detects an idler photon after a filter having line width $\Gamma$, then the signal field is put into a mixed state of the form[†]

$$\rho = \sum_i P_\Gamma(\omega_i) \left|\Psi_S(\omega_i)\right\rangle \left\langle\Psi_S(\omega_i)\right| \quad , \tag{6}$$

where $P_\Gamma(\omega_i)$ is determined by the filter transmission function, and

$$\left|\Psi_S(\omega_i)\right\rangle = \int d\omega\, C(\omega,\omega_i) \left|\omega\right\rangle_S \quad . \tag{7}$$

---

[†] Different states in this decomposition are not necessarily orthogonal, and therefore Eq.(6) should not be considered as a diagonalization of $\rho$.

This is one of many possible pure-state wave packets that the signal may be created in, given a trigger count at the idler detector. In order to reduce the distribution to a single pure-state packet, one must reduce the filter line width to zero, leading, in this limit, to zero efficiency for packet production.

The goal of this study is to design a PDC system that does not suffer this trade-off between wave-packet purity and packet production rate. This is accomplished by modifying the PDC process by using a microcavity to engineer the density of states of the optical field. This is analogous to cavity engineering for controlling spontaneous emission from atoms or quantum dots [10]. A study of PDC in a macroscopic cavity has been discussed in [11], but with different goals. Also, an alternative approach to engineering pure-state packets by using PDC in dispersion-tailored crystals has been studied [9, 12].

We propose a system that is based on PDC in a distributed optical microcavity. The need for a distributed cavity structure arises from the tradeoff between the phase-matching bandwidth of the PDC process and the free spectral range (FSR) of a cavity. Typically, in a bulk nonlinear-optical crystal of length $L$, the phase-matching bandwidth equals approximately $10\,c/L$ ($c =$ speed of light). A cavity with length $L_{cav}$, has FSR equal to $c/2L_{cav}$. Therefore, if one simply places such a crystal inside a cavity made of two mirrors separated by distance $L_{cav}$, then many cavity modes can be excited in the process, leading to a broad-band spectrum, and the difficulties reviewed above.

In contrast, in our proposal the cavity mirrors are integrated into the nonlinear optical crystal in the form of distributed Bragg reflectors (DBRs), created by a small, periodic modulation of the linear refractive index along the cavity axis (preferably confined by a waveguide). This is shown in Fig.2.

In this configuration, the cavity (a small ~ 0.1 *mm* gap in the center of the DBR) has a much shorter length than does the nonlinear medium ($L \simeq 4$ *mm*). In this case, only a single cavity mode falls within the phase-matching bandwidth of the PDC process, leading to a single frequency at which PDC can take place. This leads to pure-state creation of a unique pair of packets – one for the signal and one for the idler.

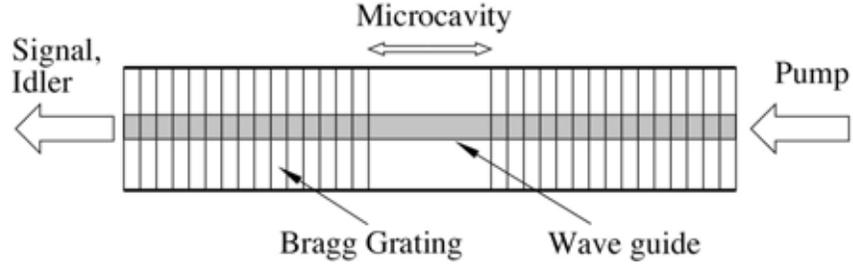

Fig.2 Microcavity consisting of a nonlinear-optical medium, with space-and-frequency-dependent linear electric susceptibility $\varepsilon(x,\omega)$, creating two DBRs (gratings) and a waveguide.

Mathematically, this means that the amplitude function factors into a product, $C(\omega,\omega') = \delta \, \psi_0(\omega) \times \phi_0(\omega')$, where $\delta$ is a small parameter. Consequently, the state in Eq.(5) simplifies to

$$| \Psi \rangle = | vac \rangle + \delta \, | 1 \rangle_{S0} \otimes | 1 \rangle_{I0} \quad , \qquad\qquad (8)$$

where the normalized single-photon packets are

$$\begin{aligned} | 1 \rangle_{S0} &= \int d\omega \; \psi_0(\omega) \; | \omega \rangle_S \\ | 1 \rangle_{I0} &= \int d\omega \; \phi_0(\omega) \; | \omega \rangle_I \end{aligned} \qquad\qquad (9)$$

This state, which displays zero entanglement within the two-photon subspace, has the benefit that no filter need be placed in front of the idler trigger detector. If *any* idler photon is counted at this detector, then it guarantees that the signal field is in a known, localized, pure-state packet. In contrast, the state in Eq.(5) typically will not factor as in Eq.(8).

To determine the state of the down-conversion field in our microcavity system, we first solve the operator form of Maxwell's equations for the electric field, and then use a quantum scattering formalism to convert the operator solution into a solution for the Schrodinger-picture quantum state, of the form Eq.(5). We then use the method of the Schmidt-mode decomposition to characterize the degree of entanglement between signal and idler fields [13].

## II – Quantum field in inhomogeneous, dispersive dielectric

A rigorous, Lagrangian-based quantum theory of light in a dispersive, spatially nonuniform and nonlinear optical dielectric is not easily developed [14]. A rigorous theory requires inclusion of medium absorption as well as dispersion in order to maintain causality [15, 16]. Instead of a rigorous approach, we use a phenomenological approach introduced by Loudon, which is approximately valid for one-dimensional light propagation in a frequency range in which the medium is transparent, and the dispersion is weak and monotonic (one-to-one relation between frequency and propagation constant) [17]. In Appendix A we present a generalization of Loudon's method to the case of an electric permittivity that varies spatially along the direction of wave propagation.

The one-dimensional, quantized-field wave equation, for a single field polarization in the presence of a space-and-frequency-dependent linear electric susceptibility $\varepsilon(x,\omega)$ and a source term $\hat{J}(x,t)$, can be written in the frequency domain as

$$\left[ \partial_x^2 + \varepsilon(x,\omega)\omega^2 / c^2 \right] \tilde{E}(x,\omega) = \tilde{J}(x,\omega) \quad , \tag{10}$$

with $\varepsilon_0$ being the vacuum permittivity (in SI units). The orthogonality properties of the modes are simplified by assuming that the (dimensionless) dielectric permittivity $\varepsilon(x,\omega)$ factors into a spatial part and a spectral part: $\varepsilon(x,\omega) = \varepsilon(x) n^2(\omega)$, where the (mean) refractive index is $n(\omega)$ and the spatial part varies by small amounts $\delta\varepsilon(x)$ around unity: $\varepsilon(x) = 1 + \delta\varepsilon(x)$. An equation of the type Eq.(10) may be written for each of the two field polarizations. The Fourier-transformed field and source operators are

$$\tilde{E}(x,\omega) = \int_{-\infty}^{\infty} dt \, e^{i\omega t} \hat{E}(x,t) \quad , \qquad \tilde{J}(x,\omega) = \int_{-\infty}^{\infty} dt \, e^{i\omega t} \hat{J}(x,t) \, . \tag{11}$$

$\hat{E}(x,t)$ is the Hermitian electric-field operator, $\hat{E}(x,t) = \hat{E}^{(+)}(x,t) + \hat{E}^{(-)}(x,t)$, where the positive-frequency part is (Appendix A)[✝]

$$\hat{E}^{(+)}(x,t) = i\int_0^\infty d\omega\, l(\omega)\, \hat{a}(\omega) u(x,\omega) e^{-i\omega t} \quad , \tag{12}$$

and $l(\omega) = \sqrt{\hbar\omega / 2\varepsilon_0 S}$. The effective beam cross sectional area is $S$. The monochromatic mode functions are denoted $u(x,\omega)$. The annihilation operators commute according to

$$[\hat{a}(\omega), \hat{a}^\dagger(\omega')] = \delta(\omega - \omega'). \tag{13}$$

Equations (11) and (12) imply the relation $\tilde{E}(x,\omega) = i2\pi\, l(\omega)\, \hat{a}(\omega) u(x,\omega)$.

The modes obey the homogeneous equation,

$$\left[\partial_x^2 + \varepsilon(x,\omega)\omega^2 / c^2\right]u(x,\omega) = 0 \quad . \tag{14}$$

The modes are orthogonal according to,

$$n^2(\omega)\int_{-\infty}^\infty dx\, \varepsilon(x) u(x,\omega) u*(x,\omega') = \delta(\omega - \omega'), \tag{15}$$

and complete,

$$\varepsilon(x)\int_{-\infty}^\infty d\omega\, n^2(\omega) u(x,\omega) u*(x',\omega) = \delta(x - x'). \tag{16}$$

---

[✝] To this field there can be added another term, $i\int_0^\infty d\omega\, l(\omega)\, \hat{b}(\omega) v(x,\omega) e^{-i\omega t}$, where the $v(x,\omega)$ are a set of modes that are linearly independent of $u(x,\omega)$, and $\hat{b}(\omega)$ is a boson operator that commutes with $\hat{a}(\omega)$.

The primary differences between this formulation and the free-space one are the forms of the mode functions and the inclusion of the weight factor $n^2(x,\omega)$ in the orthogonality integral.

The solution of Eq.(10) can be written (see Appendix B)

$$\tilde{E}(x,\omega) = \tilde{E}_{free}(x,\omega) + \int_0^L dx' \, \tilde{K}(x,x',\omega)\tilde{J}(x',\omega) \qquad , \qquad (17)$$

where $L$ is the length of the source medium, and $\tilde{E}_{free}(x,\omega) = i2\pi \, l(\omega)\hat{a}_{free}(\omega)u(x,\omega)$ is the homogeneous solution representing the input fields, which for our purposes will be assumed to be in the vacuum state.

We are concerned with the case that a pump laser pulse is incident on the cavity, and the outgoing pulse of PDC photons is detected far outside the cavity, at a much later time. We show in Appendix B that, for this case, the Green function is well approximated as

$$\tilde{K}(x,x',\omega) = \frac{-i\pi c^2}{\omega}u(x,\omega)u*(x',\omega) . \qquad (18)$$

Explicit forms for the mode functions are given below, for a particular cavity geometry. Combining Eqs.(18), (17), and (12) shows that the annihilation operator in the long-time, far-field limit, is given by

$$\hat{a}(\omega) = \hat{a}_{free}(\omega) - \frac{c^2}{2\omega \, l(\omega)}\int_0^L dx' \, u*(x',\omega)\tilde{J}(x',\omega) \quad . \qquad (19)$$

## III – Solutions for signal and idler field operators

The analysis in the previous section must be performed for both the signal and idler fields, $\tilde{E}_S(x,\omega)$ and $\tilde{E}_I(x,\omega)$, which obey

$$\left[\partial_x^2 + \varepsilon_S(x,\omega)\omega^2 / c^2\right]\tilde{E}_S(x,\omega) = \tilde{J}_S(x,\omega)$$
$$\left[\partial_x^2 + \varepsilon_I(x,\omega)\omega^2 / c^2\right]\tilde{E}_I(x,\omega) = \tilde{J}_I(x,\omega)$$

(20)

where the relative permittivities $\varepsilon_S(x,\omega)$, $\varepsilon_I(x,\omega)$ are in general different because of birefringence and differing polarizations. The field operators are expanded in their own sets of modes,

$$\hat{E}_S^{(+)}(x,t) = i\int_0^\infty d\omega\, l(\omega)\,\hat{a}_S(\omega)u_S(x,\omega)e^{-i\omega t}$$
$$\hat{E}_I^{(+)}(x,t) = i\int_0^\infty d\omega\, l(\omega)\,\hat{a}_I(\omega)u_I(x,\omega)e^{-i\omega t}$$

(21)

The signal modes $u_S(x,\omega)$ and idler modes $u_I(x,\omega)$ obey, respectively,

$$\left[\partial_x^2 + \varepsilon_S(x,\omega)\omega^2 / c^2\right]u_S(x,\omega) = 0$$
$$\left[\partial_x^2 + \varepsilon_I(x,\omega)\omega^2 / c^2\right]u_I(x,\omega) = 0$$

(22)

The signal modes are mutually orthogonal as in Eq.(15), as are the idler modes. But signal modes are not necessarily spatially orthogonal to idler modes.

Since we are concerned with spontaneous down conversion, we can treat the sources using perturbation theory, in which the source is given by zeroth-order initial conditions. The source terms are, in the time domain,

$$\hat{J}_S(x,t) = \chi(x)E_p^{(+)}(x,t)\hat{E}_{I,free}^{(-)}(x,t)$$
$$\hat{J}_I(x,t) = \chi(x)E_p^{(+)}(x,t)\hat{E}_{S,free}^{(-)}(x,t)$$

(23)

where $\hat{E}_{S,free}^{(-)}$ and $\hat{E}_{I,free}^{(-)}$ are zeroth-order perturbative solutions, i.e., the free field operators in the absence of any interaction with a pump field. The pump field operator has been replaced by a real function $E_p^{(+)}(x,t)$, under the assumption that the pump is described by a strong coherent state. We made the simplifying assumption that $\chi(x)$, which is proportional to the third-order nonlinear optical susceptibility, is frequency independent. The pump is

$$E_P^{(+)}(x,t) = (2\pi)^{-1} \int d\omega \, E_P(\omega) e^{-i\omega t} \, w(x,\omega) \quad , \tag{24}$$

where $w(x,\omega)$ is the pump mode and $E_P(\omega)$ is the pump amplitude spectrum. Therefore, in the frequency domain the source terms are

$$\tilde{J}_S(x,\omega) = -i\chi(x) \int d\omega' \, E_P(\omega+\omega') w(x,\omega+\omega') l(\omega') \hat{a}_I^{\dagger}(\omega') u_I *(x,\omega')$$
$$\tilde{J}_I(x,\omega) = -i\chi(x) \int d\omega' \, E_P(\omega+\omega') w(x,\omega+\omega') l(\omega') \hat{a}_S^{\dagger}(\omega') u_S *(x,\omega') \tag{25}$$

The signal field operator is then

$$\tilde{E}_S(x,\omega) = \tilde{E}_{S,free}(x,\omega) + \int_0^L dx' \, \tilde{K}_S(x,x',\omega) \tilde{J}_S(x',\omega) \quad , \tag{26}$$

where

$$\tilde{K}_S(x,x',\omega) = \frac{-i\pi c^2}{\omega} u_S(x,\omega) u_S *(x',\omega) . \tag{27}$$

Likewise the idler field is given by

$$\tilde{E}_I(x,\omega) = \tilde{E}_{I,free}(x,\omega) + \int_0^L dx' \, \tilde{K}_I(x,x',\omega) \tilde{J}_I(x',\omega)$$
$$\tilde{K}_I(x,x',\omega) = \frac{-i\pi c^2}{\omega} u_I(x,\omega) u_I *(x',\omega) \tag{28}$$

The output photon annihilation operator is, from Eq.(19),

$$\hat{a}_S(\omega) = \hat{a}_{S,free}(\omega) - \frac{c^2}{2\omega l(\omega)} \int_0^L dx' \, u_S *(x',\omega) \tilde{J}_S(x',\omega) \quad , \tag{29}$$

which can be written as

$$\hat{a}_S(\omega) = \hat{a}_{S,free}(\omega) + \int_0^\infty d\omega' \, B(\omega,\omega')\hat{a}^\dagger_{I,free}(\omega') \quad , \tag{30}$$

where

$$B(\omega,\omega') = \frac{i c^2 l(\omega')}{2\omega l(\omega)} E_P(\omega + \omega') \int_0^L dx \, \chi(x) \, w(x,\omega + \omega') u_S *(x',\omega) u_I *(x,\omega') \quad . \tag{31}$$

This amplitude function is a product of the pump spectrum, evaluated at the sum of the signal and idler frequencies, and the phase-matching integral involving the modes of the pump, signal, and idler. We will assume that $\chi(x)$ is independent of $x$.

Likewise, the idler annihilation operator is given by

$$\hat{a}_I(\omega) = \hat{a}_{I,free}(\omega) + \int_0^\infty d\omega' \, B(\omega',\omega)\hat{a}^\dagger_{S,free}(\omega') \quad , \tag{32}$$

Notice the interchange of the arguments of $B$ compared to in Eq.(30). Equations (30) and (32) comprise a multimode Bogliubov transformation, describing weak-field quadrature squeezing of the signal and idler fields.

## III – From quantum fields to wave functions

The solutions (Eqs. (30) and (32)) are in the Heisenberg picture, where the operators evolve and the states are constant. To study entanglement in a two-photon quantum state of the form in Eq.(5), we need to convert these solutions to the Schroedinger picture. In order to make a connection between these pictures, we view the calculation as a scattering theory, or equivalently, an example of input-output theory [18]. We are not concerned with the detailed dynamics occurring inside the nonlinear medium, but rather the properties of the scattered field after it and the pump pulse have exited the medium. The Green function (18) was constructed for this case.

To convert between pictures, it is sufficient to note that the scattering solutions for the operators (Eqs. (30) and (32)) can be generated by a unitary transformation $\hat{U} = e^{-i\hat{H}}$, where $\hat{H}$ is a Hermitian operator, which plays a role similar to a dimensionless interaction Hamiltonian operator. In the weak-scattering limit (Born approximation) being considered here, the transformation is approximated by $\hat{U} \approx (1 - i\hat{H})$. In the Heisenberg picture, an operator $\hat{O}_0$ evolves under the action of $\hat{U}$ as

$$\hat{O} = \hat{U}^{-1} \hat{O}_0 \hat{U} \approx \hat{O}_0 + i[\hat{H}, \hat{O}_0] \tag{33}$$

It is easy to see that the interaction Hamiltonian that generates the output (scattered) electric field operators $\hat{a}_S(\omega)$ and $\hat{a}_I(\omega)$ in Eqs.(30) and (32) has the form

$$\hat{H} = i \int_0^\infty d\omega_1 \int_0^\infty d\omega_2 \ B(\omega_1, \omega_2) \hat{a}_S^\dagger(\omega_1) \hat{a}_I^\dagger(\omega_2) + h.c.. \tag{34}$$

Using this Hamiltonian in Eq.(33), and the commutation relation $[\hat{a}_i(\omega), \hat{a}_j^\dagger(\omega')] = \delta_{ij} \delta(\omega - \omega')$ (with $i, j=S,I$) yields results identical to those in Eqs.(30) and (32). This verifies the correctness of the Hamiltonian in Eq.(34).

The state evolution in the Schroedinger picture is then given simply by

$$\begin{aligned} |\psi\rangle &= e^{-i\hat{H}} |vac\rangle \approx (1 - i\hat{H}) |vac\rangle \\ &= |vac\rangle + \int_0^\infty d\omega_1 \int_0^\infty d\omega_2 \ B(\omega_1, \omega_2) |\omega_1\rangle_S |\omega_2\rangle_I \end{aligned}, \tag{35}$$

where $|\omega_1\rangle_S |\omega_2\rangle_I$ is the state a pair of photons at frequencies $(\omega_1, \omega_2)$. Comparing this to Eq.(5) shows that the two-photon amplitude is $C(\omega, \omega') = B(\omega, \omega')$, where $B(\omega, \omega')$ is given in Eq.(31).

## IV – Quantifying entanglement using the Schmidt-mode decomposition

As pointed out in the context of PDC by Law et. al. [13], the degree of entanglement present in the two-photon state (35) can best be understood using a discrete basis comprised of Schmidt modes. Schmidt modes are a set of perfectly correlated pairs of pure-state, spatial-temporal wave packets.

The Schmidt-mode decomposition of a bipartite quantum state allows a continuous double integral over state parameters, such as $(\omega, \omega')$ in Eq.(35) or (5), to be rewritten as a single summation over discrete mode labels. This is done by making a diagonal decomposition of the amplitude function [13],

$$B(\omega, \omega') = \sum_{j=0}^{\infty} \sqrt{\lambda_j}\ \psi_j(\omega)\ \phi_j(\omega')\ ,$$  (36)

where the coefficients $\lambda_j$ and the mode functions $\psi_j(\omega), \phi_j(\omega')$ are to be determined. The coefficients obey $\sum_j \lambda_j = 1$. The mode functions are orthonormal and complete (see below),

$$\int \psi_j *(\omega)\ \psi_k(\omega) d\omega = \delta_{jk}, \quad \int \phi_j *(\omega)\ \phi_k(\omega) d\omega = \delta_{jk}\ ,$$  (37)

$$\sum_{j=0}^{\infty} \psi_j *(\omega)\ \psi_j(\omega') = \delta(\omega - \omega'), \quad \sum_{j=0}^{\infty} \phi_j *(\omega)\ \phi_j(\omega') = \delta(\omega - \omega')\ .$$  (38)

Such a decomposition is possible for any well behaved function $B(\omega, \omega')$. This leads immediately to a diagonal-summation form for the state,

$$|\psi\rangle = |vac\rangle + \sum_{j=0}^{\infty} \sqrt{\lambda_j}\ |1\rangle_{S,j} \otimes |1\rangle_{I,j}\ ,$$  (39)

where the orthonormal, single-photon wave-packet states are given by

$$|1\rangle_{Sj} = \int_0^\infty d\omega\, \psi_j(\omega)\,|\,\omega\,\rangle_S$$
$$|1\rangle_{Ij} = \int d\omega\, \phi_j(\omega)\,|\,\omega\,\rangle_I$$

(40)

Each $j$-value identifies a pair of perfectly correlated single-photon states of the signal and idler fields. The operator $\hat{b}_{Sj}^\dagger \otimes \hat{b}_{Ij}^\dagger$ creates a photon pair in a particular wave-packet pair, that is, $|1\rangle_{Sj} = \hat{b}_{Sj}^\dagger|vac\rangle, |1\rangle_{Ij} = \hat{b}_{Ij}^\dagger|vac\rangle$. These creation operators are

$$\hat{b}_{Sj}^\dagger = \int d\omega\, \psi^*_j(\omega)\, \hat{a}_S^\dagger(\omega)$$
$$\hat{b}_{Ij}^\dagger = \int d\omega\, \phi^*_j(\omega)\, \hat{a}_I^\dagger(\omega)$$

(41)

The electric field operators (12) can be written, using Eq.(38), in terms of the wave-packet annihilation operators and two sets of wave-packets modes $\{v_{Sj}(x,\omega)\}, \{v_{Ij}(x,\omega)\}$ as

$$\hat{E}_S^{(+)}(x,t) = i\sum_{j=0}^\infty \hat{b}_{Sj}\, v_{Sj}(x,\omega), \quad \hat{E}_I^{(+)}(x,t) = i\sum_{j=0}^\infty \hat{b}_{Ij}\, v_{Ij}(x,\omega) \quad.$$

(42)

The spatial-temporal wave-packet modes of the signal or idler field are, respectively,

$$v_{Sj}(x,t) = \int_0^\infty d\omega\, l(\omega)\psi_j(\omega)u_S(x,\omega)\exp[-i\omega t]$$
$$v_{Ij}(x,t) = \int_0^\infty d\omega\, l(\omega)\phi_j(\omega)u_I(x,\omega)\exp[-i\omega t]$$

(43)

If not for the $l(\omega) = \sqrt{\hbar\omega/2\varepsilon_0 S}$ factor in the integrands, the signal wave-packet modes would form a spatially orthogonal set. Likewise for the idler modes. In cases where the spectral widths of the modes are small compared to the central frequencies, the modes are nearly orthogonal. In free space, where detection takes place, the signal wave-packet modes are given more simply by

$$v_{Sj}(x,t) = \int_0^\infty d\omega \; \psi_j(\omega) \; \frac{\exp[-i\omega(t - x/c_0)]}{\sqrt{D}} , \qquad (44)$$

where $D$ is a quantization length. A similar form holds for the idler.

To determine the forms of the Schmidt states and wave packets, define the functions $\rho_S(\omega_1, \omega_2)$ and $\rho_I(\omega_1, \omega_2)$ as follows, and use Eqs.(36) and (37) as an ansatz. Then

$$\rho_S(\omega_1, \omega_2) = \int d\omega B(\omega_1, \omega) B*(\omega_2, \omega)$$
$$= \sum_j \lambda_j \psi_j(\omega_1) \psi*_j(\omega_2) \qquad . \qquad (45)$$

$$\rho_I(\omega_1, \omega_2) = \int d\omega B(\omega, \omega_1) B*(\omega, \omega_2)$$
$$= \sum_j \lambda_j \phi_j(\omega_1) \phi*_j(\omega_2) \qquad . \qquad (46)$$

Orthogonality then gives two eigenvalue problems,

$$\int \rho_S(\omega, \omega') \psi_j(\omega') d\omega' = \lambda_j \psi_j(\omega)$$
$$\int \rho_I(\omega, \omega') \phi_j(\omega') d\omega'' = \lambda_j \phi_j(\omega) \qquad . \qquad (47)$$

Because the functions $\rho_S(\omega_1, \omega_2)$ and $\rho_I(\omega_1, \omega_2)$ are Hermitian, the eigenvalues are real and the eigenfunctions are orthogonal and complete.

The functions $\rho_S(\omega_1, \omega_2)$ and $\rho_I(\omega_1, \omega_2)$ represent the reduced density operators for signal and idler fields, if the other is traced over. This corresponds to, for example, the state of the signal field after an idler photon is detected in the absence of any spectral filtering. Since the frequency of the idler photon is undetermined, the signal photon is put into a conditional state described by the density operator,

$$\hat{\rho}_S = Tr_I(|\psi\rangle\langle\psi|) = \int d\omega \,_S\langle\omega|\psi\rangle\langle\psi|\omega\rangle_S$$
$$= \int d\omega' \int d\omega'' \rho_S(\omega', \omega'') |\omega'\rangle\langle\omega''| \qquad , \qquad (48)$$

where $\rho_S(\omega',\omega'')$ is given by Eq.(45). Likewise, for the reduced density operator of the idler following a detection of the signal photon,

$$
\begin{aligned}
\hat{\rho}_I = Tr_S\left(|\psi\rangle\langle\psi|\right) &= \int d\omega_I \langle\omega|\psi\rangle\langle\psi|\omega\rangle_I \\
&= \int d\omega' \int d\omega'' \rho_I(\omega',\omega'')|\omega'\rangle\langle\omega''|
\end{aligned}
\qquad (49)
$$

where $\rho_I(\omega',\omega'')$ is given by Eq.(46). Thus, $\rho_S(\omega',\omega'')$ and $\rho_I(\omega',\omega'')$ are density matrices in the single-photon-frequency (energy) basis.

These reduced density operators can also be written as

$$
\begin{aligned}
\hat{\rho}_S &= \sum_{j=0}^{\infty} \lambda_j |1\rangle_{Sj}\ {}_{Sj}\langle 1| \\
\hat{\rho}_I &= \sum_{j=0}^{\infty} \lambda_j |1\rangle_{Ij}\ {}_{Ij}\langle 1|
\end{aligned}
\qquad (50)
$$

This illustrates that our goal must be to engineer the microcavity mode density such that only a single eigenvalue $\lambda_j$ is nonzero. This will occur if $B(\omega,\omega')$, given by Eq.(31), is proportional to a product, $\psi_0(\omega) \times \phi_0(\omega')$, for some functions $\psi_0(\omega)$ and $\phi_0(\omega)$. Then, following detection of an idler photon (without any filters), the signal field will be known to contain a single photon is a known spatial-temporal wave-packet mode.

If more than one eigenvalue is nonzero, then useful characterizations of the state are given by the entropy of entanglement [13],

$$
S = -\sum_{j=1}^{\infty} \lambda_j \log_2 \lambda_j \quad, \qquad (51)
$$

the purity parameter, $p$, and the cooperativity number, $K$, [19]

$$
p = Tr(\hat{\rho}_S^2), \quad K = \left(\sum_{j=0}^{\infty} \lambda_j^2\right)^{-1} \quad. \qquad (52)
$$

$K$ estimates the number of populated Schmidt modes. The purity and the cooperativity number are related by $p = 1/K$ [12].

### V – Microcavity modes

Here we propose and analyze a specific microcavity geometry that controls the density of electromagnetic modes in a way that leads to a nearly factorized two-photon state (in addition to a vacuum component). For the calculation, we use a simplified model, shown in Fig.3. In the design, one of the DBRs is replaced by a wide-band, planar mirror, separated from the remaining DBR by an air gap of length $d$. The mirror amplitude reflectivity and transmissivity are real and equal to $\rho$ and $\tau$ for light incident from the right ($x > 0$), and $-\rho$ and $\tau$ for light incident from the left ($x < 0$). These obey $\rho^2 + \tau^2 = 1$. We assume that these values are constant over a frequency interval wider than any being considered here. The remaining DBR, which is embedded in the second-order nonlinear medium of length $L$, has linear reflectivity and transmissivity coefficients $r', t'$ for light incident from the right, and $r, t$, for light incident from the left. Explicit expressions for the DBR coefficients are given below. The labeling of the field amplitudes in the three distinct regions (1, 2, and 3) follows the notation used below to analyze the DBR.

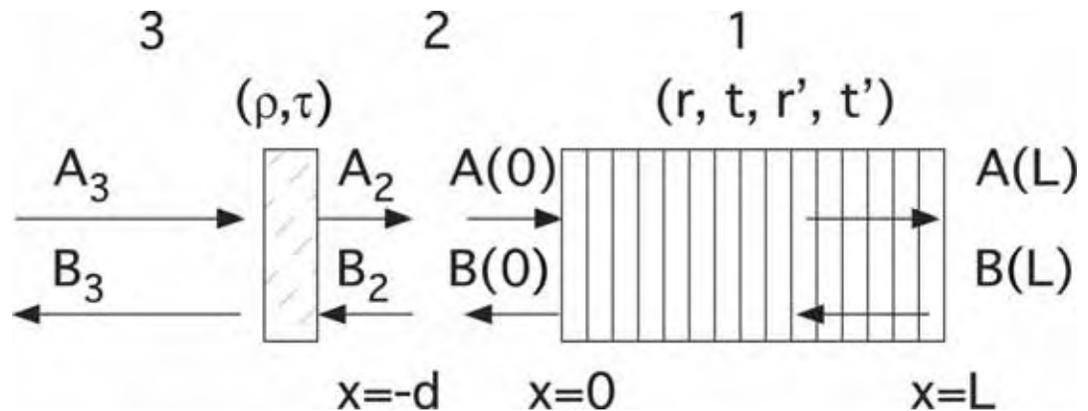

Fig.3 Microcavity with one DBR occupying $x = [0, L]$, and a mirror at $x = -d$, separated by an air gap.

The mode functions obey Eq.(14), with the linear permittivity $\varepsilon(x)n^2(\omega)$ chosen to be appropriate for the signal, idler, or pump. Within the DBR, the spatial part equals a constant plus a small periodic part,

$$\varepsilon(x) = 1 + \Delta\varepsilon\cos(Kx), \tag{53}$$

where $K = 2\pi/\Lambda$ is the spatial frequency of the grating with period $\Lambda$. We write the mode as a sum of right- and left-going waves,

$$u(x,\omega) = A(x,\omega)e^{ik(\omega)x} + B(x,\omega)e^{-ik(\omega)x}, \tag{54}$$

where $k(\omega) = n(\omega)\omega/c$. Using the standard coupled-mode approximation [20] leads to

$$\begin{aligned}\partial_x A(x,\omega) &= i\kappa\, B(x,\omega)e^{i\Delta(\omega)x} \\ \partial_x B(x,\omega) &= -i\kappa\, A(x,\omega)e^{-i\Delta(\omega)x}\end{aligned}, \tag{55}$$

where the coupling constant is $\kappa(\omega) = \Delta\varepsilon\,\omega^2/(4k(\omega)c^2) = (\Delta n/2)k(\omega)$, where $\Delta n$ is the modulation of the refractive index. In the following we will neglect the frequency dependence of $\kappa(\omega)$. The detuning parameter is $\Delta(\omega) = K - 2k(\omega)$. The boundary values $A(0,\omega)$, $B(L,\omega)$ are fixed by the normalization condition Eq.(15). This amounts to taking $A_3$ to be the free-field monochromatic mode function $(2\pi cn)^{-1/2}\,e^{ikx}$ at $x = -d$. Equation (55) is easily solved, and the solution inside the DBR, given in **Appendix C**, has the form,

$$\begin{aligned}A(x,\omega) &= Q(x,\omega)A(0,\omega) + i\kappa\, P(x,\omega)B(L,\omega)e^{i\Delta(\omega)L} \\ B(x,\omega) &= -i\kappa\, W(x,\omega)A(0,\omega) + V(x,\omega)B(L,\omega)\end{aligned}, \tag{56}$$

where the functions $Q(x,\omega)$, $P(x,\omega), W(x,\omega), V(x,\omega)$ are given explicitly in **Appendix C**.

As a whole, the DBR acts as a beam splitter, with the input-output relations,

$$\begin{bmatrix} A(L) \\ B(0) \end{bmatrix} = \begin{bmatrix} t & r' \\ r & t' \end{bmatrix} \begin{bmatrix} A(0) \\ B(L) \end{bmatrix}. \tag{57}$$

The matrix in this transformation is unitary, as required by the Stokes' relations:

$|t| = |t'|, |r| = |r'|, \ rt* + r'*t' = 0$ [8]. Specifically, the reflectivities and transmissivities are found

to be

$$\begin{aligned} t &= t' = Q(L) \\ r &= r'e^{-i\Delta(\omega)L} = -i\kappa W(0) \end{aligned} \tag{58}$$

Figure 4 shows the reflectivity $|r|^2$ of the DBR (only) vs input frequency, measured in $k$ (in units mm$^{-1}$).

    This is now a standard input/output calculation for a two-mirror cavity. Solutions are

given in **Appendix C**. Illustrations of the results are shown in Fig.4. For an input, $A_3$, incident

only from the left, Fig.4(b) shows the cavity reflectivity $|R|^2$. When the scale is expanded, as in

(c), the narrow cavity dip is clear. Fig.4(d) shows the intracavity intensity $|A_2|^2$.

    We consider here only the case that the DBR is sufficiently long that no light can

penetrate through it in the spectral region of interest. We will therefore neglect any (vacuum)

input, $B(L)$, from the right side of the cavity. This is an approximation, which is valid only for

the frequencies inside the DBR stop band. Outside the stop band, light can penetrate into the

DBR from the right. As we will see below, the mode of interest is concentrated within the stop band.

    The mode functions in the DBR are proportional to $E_1(x)$, with the frequency

dependence restored in the notation:

$$u(x,k) = \frac{\tau}{f(k)} \Big[ Q(x,k)e^{ikx} - i\kappa W(x,k)e^{-ikx} \Big] \ , \tag{59}$$

where $k = k(\omega)$. Normalization of the modes are dominated by the integral to infinity outside the cavity. Therefore, $A_3$ is the free-field monochromatic field operator.

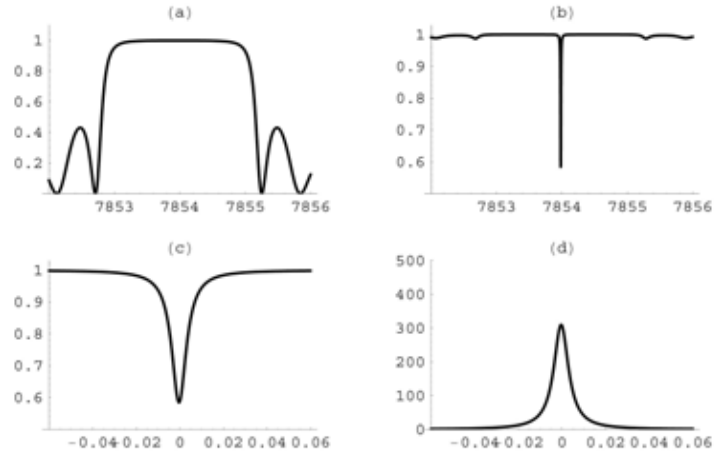

Fig.4 – (a) Reflectivity $|r|^2$ of DBR (only) vs input frequency, measured in $k$ (in units mm$^{-1}$), (b) Cavity reflectivity $|R|^2$; (c) Cavity reflectivity $|R|^2$ zoomed in, (d) Intracavity intensity $|A_2|^2$. (Parameter values: $L = 4$ mm$^{-1}$, $d = 0.1999$ mm, $\kappa = 1$ mm$^{-1}$, $K = 4\pi$ /(0.8×10$^{-3}$) mm$^{-1}$, $\rho = \sqrt{0.99}$, center wavelength $\lambda = 800$ nm, background refractive index = 1)

## VI – Schmidt wave-packet modes

The mode functions found in the previous section are used with (31) to calculate the two-photon amplitude $B(\omega, \omega')$. In order to control the mode density at both the idler and signal frequencies, we assume that a DBR grating is created having two spatial periods – one with spatial frequency $K_S = 4\pi n_S / \lambda_S$ that causes reflection of signal waves and the other with $K_I = 4\pi n_I / \lambda_I$ that reflects idler waves: $\varepsilon(x) = 1 + \Delta\varepsilon[\cos(K_S x) + \cos(K_I x)]$. The grating coupling strength of each grating component is chosen as $\kappa = 2$ mm$^{-1}$, corresponding to $\Delta n / n = 5\times10^{-4}$.

We consider a model based on the material parameters of KTP crystal, with optical axis at an angle 0.739095 rad to the $k$-vector of the e-wave, for type-II bulk phase matching. The signal and idler fields are phase matched at the degenerate center wavelength (in air) $\lambda_S = \lambda_I = 800$ nm. The pump wavelength is 400 nm. Sellmeier formulas are used for the (first-order) dispersion of ordinary (idler) and extraordinary (pump and signal) waves. The pump is assumed to propagate only in the $+x$ direction, and suffer no reflections at the DBR. This is reasonable, as dispersion puts the pump $k$-value outside of the second-order reflection band gaps. We assume a transform-limited pump-pulse spectrum $E_P(\omega) = E_{P0} \exp(-\omega^2/\sigma^2)$, with $\sigma = 0.3 \times 10^{12} rad/s$. The duration of such a pulse is about 10 ps. The cavity air gap again has length $d = 0.1999$ mm, in order to resonate the signal and idler. The crystal length $L$ is 4 mm, giving $\kappa L = 8$, This means that the modes decay from the cavity air gap into the crystal, having very weak amplitude at the other end of the crystal; so the PDC output is almost entirely from the cavity-mirror end at $x = -d$. The refractive indices and derivatives with respect to angular frequency are: $n_I = 1.6605$; $n_S = 1.6047$; $n_P = 1.6326$; $k'_I = 5.6149 \times 10^{-9}$; $k'_S = 5.4212 \times 10^{-9}$; $k'_P = 5.6949 \times 10^{-9}$.

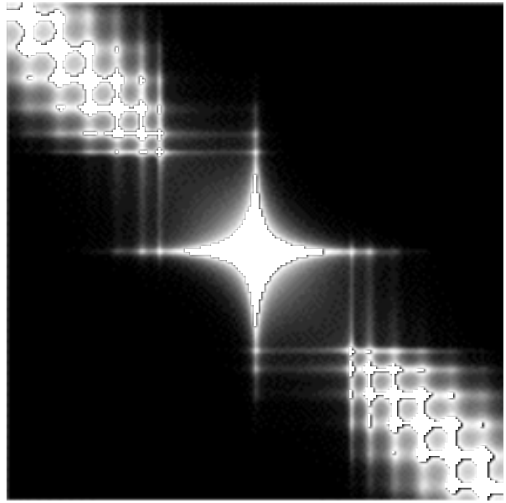

Fig.5. Absolute value of two-photon amplitude $|B(\omega, \omega')|$, for $\rho = \sqrt{0.95}$. The horizontal (vertical) axis is signal (idler) frequency, in a range $\{2.3552, 2.3572\} \times 10^{15}$ rad/sec. The axes are labeled by grid point number.

The two-photon amplitude for external mirror amplitude reflectivity $\rho = \sqrt{0.95}$ is shown in Fig.5. The bright cross-shaped peak at the center is the double-cavity mode at the center of the two DBR band gaps (a wide, cross-shaped region centered at the figure center). The weak, unwanted amplitudes are in the lower right and upper left regions; these are allowed by phase matching and pump spectrum, but are outside of the cavity-controlled central region. Two questions arise: how dominant is the central cross-shaped peak, and is it factorizable? The Schmidt-mode calculation answers these.

For the numerical solution of Eqs.(47) we used a two-dimensional grid of up to $1191 \times 1191$ points in a finite spectral interval, containing the DBR gap. This leads to eigenvalues given in **Table 1**.

| Schmidt-mode eigenvalues for different values of cavity-mirror reflectivity $\rho^2$ | | | | | |
|---|---|---|---|---|---|
| $\rho^2$ | j=1 | j=2 | j=3 | j=4 | j=5 |
| 0.95 | 0.951 | 0.0196 | 0.0196 | 0.0044 | 0.0044 |
| 0.99 | 0.998 | 0.0007 | 0.0007 | 0.0002 | 0.0002 |

Examples of normalized Schmidt mode functions for the signal field are shown in Fig.6, for the case of 0.95 cavity-mirror reflectivity. The idler modes look quite similar. It is seen that the cavity mode strongly dominates the first Schmidt mode (a), which contains over 95% of the probability. Other modes (b-d) have spectra that are concentrated outside of the DBR reflectivity gap.

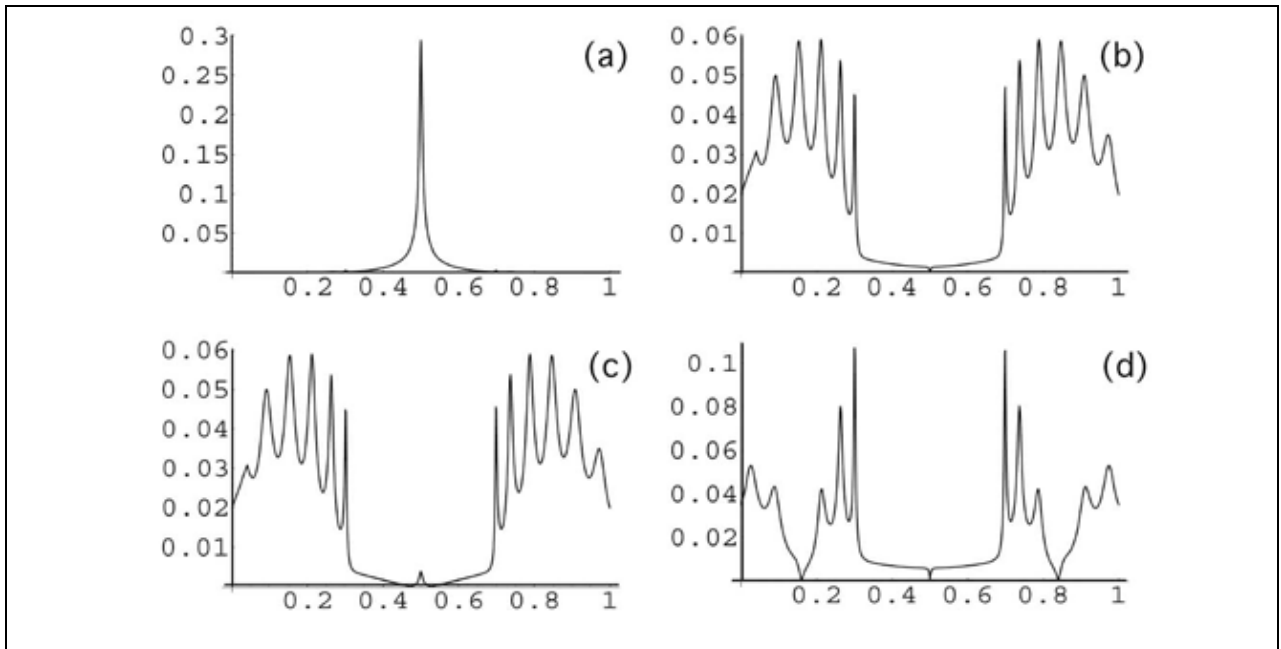

Fig.6  First four signal Schmidt-mode spectral eigenfunctions versus frequency. The horizontal axis is signal frequency, in a range {2.3552, 2.3572}$\times 10^{15}$ rad/sec. The first Schmidt mode (a) contains over 95% of the probability. Other modes (b-d) are concentrated outside the DBR reflectivity region.

## VII – Results and Conclusions

Our proposal is to engineer the density of states of the vacuum prior to the creation of idler-signal photon pairs. Spectral filtering of the field occurs prior to photon creation rather than afterward as in most other schemes. Our scheme for doing this is to fabricate a high-finesse distributed cavity throughout a second-order nonlinear crystal. The high-finesse cavity mode occupies a spectral interval much narrower than the bandwidth of the pulsed pump laser field, suppressing the spectral correlation, or entanglement, between signal and idler photons.

Our simple model explored here treats only a one-dimensional problem, under the assumption that wave-guiding structures can be used to control the transverse degrees of freedom. We treated a simplified geometry, comprised of a linear-index distributed Bragg reflector (DBR) embedded in a 4-mm long nonlinear crystal, with a small (~ 0.2 mm) air gap and adjacent planar mirror to form a cavity. In practice, the air gap would be replaced with a guiding structure to prevent losses.

For cavity-mirror reflectivity $\rho^2 = 0.99$, we find that the central peak contains over 99% of the probability for photon pair creation (see Table 1), without any external filtering. This means that if any idler photon is detected, then the signal photon will be in the first Schmidt mode with 99% probability. This does not require spectral filtering before detection. This result appears to be very promising for high-rate production of pure-state, controlled single-photon wave packets.

Other Schmidt modes have spectra that are concentrated outside of the DBR reflectivity spectral gap. This implies that spectral filtering outside the cavity, but before detection, could be used to decrease the false trigger rate (in which an idler photon would be paired with a signal photon in other than the first mode), without degrading the purity of the signal field state. This would increase the probability of heralded pure-state signal photon creation to greater than 99%, with a negligible decrease of overall photon production rate.

A more realistic model would account for the dispersion introduced by the confinement of light (waveguide dispersion). Fortunately, the increased dispersion will only serve to further increase the first Schmidt-mode probability, according to preliminary calculations

Fabricating suitable waveguide microcavity structures requires developing new techniques. The challenges are: 1. Creating linear DBRs in a bulk-phase-matched, low-loss nonlinear optical medium, and 2. Creating type-II phase matching (in which the signal and idler are orthogonally polarized) in a suitable waveguide structure. It is possible that quasi-phase-matching techniques (in which the nonlinearity is modulated with a long period) will prove superior to birefringent phase matching for this application.

An alternate device geometry may take advantage of the microcavity control while eliminating the need for type-II interactions. According to the ideas presented here, a type-I waveguided microcavity would produce weak quadrature-squeezed light with high probability to be dominated by a single controlled spatial-temporal mode. By constructing two identical such cavities, and allowing their outputs to interfere in a 50/50 superposition, twin-photon pulses will be created, having the same number of photons in each pulse. A photon in one pulse may then serve as the trigger for heralding the occurrence of the other.

**Acknowledgements**

We appreciate fruitful discussions with Peter Smith and Brian Smith about nonlinear optical waveguides, and Colin McKinstrie about quantization in dielectrics. This work was supported by the National Science Foundation, through its Information Technology Research (ITR) program, grant number PHY-0219460.

**APPENDIX A: QUANTIZATION IN A NONUNIFORM DISPERSIVE DIELECTRIC**

Here we present a generalization of Loudon's quantization method [17] to the case of a dispersive electric permittivity that varies spatially along the direction of wave propagation. The orthogonality properties of the modes are simplified by assuming that the (dimensionless) dielectric permittivity $\varepsilon(x,\omega)$ factors into a spatial part and a spectral part: $\varepsilon(x,\omega) = \varepsilon(x)n^2(\omega)$, where the (mean) refractive index is $n(\omega)$ and the spatial part varies by small amounts $\delta\varepsilon(x)$ around unity, $\varepsilon(x) = 1 + \delta\varepsilon(x)$. Maxwell's equations in the absence of free charges and currents (in S.I. units) are:

$$\nabla \times \underline{B} = \mu_0 \partial_t \underline{D}$$
$$\nabla \times \underline{E} = -\partial_t \underline{B} \tag{A1}$$

$$\nabla \cdot \underline{B} = \nabla \cdot \underline{D} = 0$$

The fields are given by the vector potential, $\underline{E} = -\partial_t \underline{A}$ , $\underline{B} = \nabla \times \underline{A}$ .

Consider the case of a plane wave propagating in the $x$ direction. The vector potential, electric field, and displacement field are polarized in the y direction, $\underline{A} = \hat{y} A_y$ , $\underline{E} = \hat{y} E_y$ , $\underline{D} = \hat{y} D_y$ . The magnetic field is polarized in the $z$ direction, $\underline{B} = \hat{z} B_z$ . Then $B_z = \partial_x A_y$ and $E_y = -\partial_t A_y$ . Maxwell's equations then imply that

$$-\partial_x B_z = \mu_0 \partial_t D_y , \qquad \partial_x E_y = -\partial_t B_z \quad . \tag{A2}$$

These imply the wave equation,

$$-\partial_x^2 A_y = \mu_0 \partial_t D_y \qquad . \tag{A3}$$

We make the usual split between positive and negative frequency components,

$$\underline{A}(x,t) = \underline{A}^{(+)}(x,t) + \underline{A}^{(-)}(x,t) , \tag{A4}$$

where $\underline{A}^{(-)}(x,t) = \left[ \underline{A}^{(+)}(x,t) \right]^\dagger$ , where the dagger represents complex (or operator) conjugation. We make the ansatz

$$A_y^{(+)}(x,t) = \int_0^\infty d\omega \frac{l(\omega)}{\omega} A(\omega,t) u(x,\omega) , \tag{A5}$$

where $l(\omega) = \sqrt{\hbar\omega / 2\varepsilon_0 S}$ , and $S$ is an effective beam cross sectional (quantization) area. The amplitudes obey the additional ansatz

$$\partial_t A(\omega,t) = -i\omega A(\omega,t) , \tag{A6}$$

with solution $A(\omega,t) = a(\omega)\exp(-i\omega t)$. The mode functions are defined to obey the homogeneous wave equation,

$$\left[\partial_x^2 + \varepsilon(x) n^2(\omega) \omega^2 / c^2\right] u(x,\omega) = 0 \qquad . \qquad (A7)$$

The forms of the other fields follow from (A5), (A6) and Maxwell's equations,

$$E_y^{(+)}(x,t) = i\int_0^\infty d\omega\, l(\omega) A(\omega,t) u(x,\omega) \ , \qquad (A8)$$

$$B_z^{(+)}(x,t) = \int_0^\infty d\omega \frac{l(\omega)}{\omega} A(\omega,t)\left(\partial_x u(x,\omega)\right). \qquad (A9)$$

$$D_y^{(+)}(x,t) = i\int_0^\infty d\omega\, l(\omega) \varepsilon_0 \varepsilon(x,\omega) A(\omega,t) u(x,\omega) \qquad . \qquad (A10)$$

Together (A5-A10) provide a solution to Maxwell's equations. Equations (A8) and (A10) imply the constitutive relation in the frequency domain, $\widetilde{D}_y(x,\omega) = \varepsilon_0 \varepsilon(x,\omega)\widetilde{E}_y(x,\omega)$.

Since (A7) is a Hermitian eigenvalue problem, with eigenvalues $n^2(\omega)\omega^2 / c^2$, the modes are orthogonal, and can be normalized according to [21]

$$n^2(\omega)\int_{-\infty}^\infty dx\, \varepsilon(x) u(x,\omega) u*(x,\omega') = \delta(\omega - \omega'). \qquad (A11)$$

The derivatives of the mode functions, which serve as modes for the *B* field in (A9), are also orthogonal, without any weight factor,

$$\int_{-\infty}^\infty dx\, \left(\partial_x u(x,\omega)\right)\left(\partial_x u*(x,\omega')\right) = \frac{\omega^2}{c^2}\delta(\omega - \omega'). \qquad (A12)$$

This can be seen by integrating (A12) by parts, and using (A7). The modes are also complete,

$$\varepsilon(x)\int_{-\infty}^{\infty}d\omega\, n^2(\omega)u(x,\omega)u*(x',\omega) = \delta(x-x') . \qquad (A13)$$

The equation of motion (A10) can be generated by a total Hamiltonian operator $H$, but a subtlety arises when dispersion is present in that $H$ cannot be interpreted as an integral over a local energy density. This is because when dispersion is present, $2E \cdot \partial_t D \neq \partial_t (E \cdot D)$. [22] We generalize and simplify Loudon's treatment [17], including the spatial dependence of $\varepsilon(x)$. Using the divergence theorem as usual, Maxwell's equation imply that the rate of change of the total energy in a volume V is given by

$$\partial_t U_V = \int_V \left( E \cdot \partial_t D + \mu_0^{-1} B \cdot \partial_t B \right) d^3 x \quad . \qquad (A14)$$

We cannot time-integrate the integrand in (A14) to obtain a spatially local energy density, but we can define $V$ to equal the whole quantization volume, and time-integrate the total energy expresssion. Consider the one-dimensional problem, where we replace $d^3 x \to S d^2 x$. Substitute (A8-10), use the orthognality relations (A11, 12), and perform a local time average, giving

$$\partial_t U_V = S\varepsilon_0 \int_0^{\infty} d\omega\, l^2(\omega)\left( A^\dagger(\omega,t) \cdot \partial_t A(\omega,t) + \partial_t A^\dagger(\omega,t) \cdot A(\omega,t) \right). \qquad (A15)$$

The integrand here is an exact differential, so we can integrate to find

$$U_V = \int_0^{\infty} d\omega\, \hbar\omega\, A^\dagger(\omega,t)A(\omega,t) = \int_0^{\infty} d\omega\, \hbar\omega\, a^\dagger(\omega)a(\omega) \quad . \qquad (A16)$$

Equation (A16) shows that $\partial_t U_V = 0$, which simply means that without interactions, total energy is conserved.

We now follow the standard procedure, setting $A(\omega,t) = 2^{-1/2}[Q(\omega,t) + iP(\omega,t)]$, where $Q(\omega,t)$ and $P(\omega,t)$ are real conjugate variables, obeying Hamilton's dynamical equations,

$$\partial_t P(\omega,t) = -\frac{\partial H}{\partial Q(\omega,t)}, \quad \partial_t Q(\omega,t) = \frac{\partial H}{\partial P(\omega,t)}. \tag{A17}$$

Equation (A17) is equivalent to Eq.(A6).

To quantize the system, replace the conjugate variables by operators, obeying $[Q(\omega,t), P(\omega',t)] = i\hbar \delta(\omega - \omega')$. This then leads to the commutator

$$[A(\omega,t), A^\dagger(\omega',t)] = [a(\omega), a^\dagger(\omega')] = \delta(\omega - \omega'), \tag{A18}$$

and the Hamiltonian is

$$H = \int_0^\infty d\omega\, \hbar\omega\, a^\dagger(\omega) a(\omega) \qquad . \tag{A19}$$

Then the Heisenberg-picture equation of motion for $A$ is

$$\partial_t A(\omega,t) = -(i/\hbar)\big[A(\omega,t), H\big] = -i\omega A(\omega,t), \tag{A20}$$

again equivalent to Eq.(A6).

The subtlety in this derivation is the fact that we have not identified a local energy density for the system composed of light and medium. The presence of dispersion implies the presence of absorption at some frequency, and this requires a different procedure than followed here. For a review of rigorous approaches, see Knoll [16], whose approach leads to results similar to our final results. Following Loudon, we expect that the approximate form here will be correct if the medium is transparent and weakly dispersive in the frequency range of interest.

## APPENDIX B: GREEN FUNCTION IN A STRUCTURED DIELECTRIC

To find the causal, outgoing-wave solution of Eq.(10), with $\varepsilon(x,\omega) = \varepsilon(x)n^2(\omega)$, we add a damping term $\sigma$:

$$\left[\partial_x^2 + \varepsilon(x,\omega)[\omega^2 + i\sigma\omega]/c^2\right]\tilde{E}(x,\omega) = \tilde{J}(x,\omega) \tag{B1}$$

The spatial dependence of the damping term is somewhat unrealistic; it will be taken to zero later. To solve (B1), expand the inhomogeneous part of the solution in terms of the complete normal modes (solutions to (A7):

$$\tilde{E}(x,\omega) = \int_{-\infty}^{\infty} C(\omega,\omega')u(x,\omega')d\omega' \tag{B2}$$

Plug into (B1), operate by $\int dx\, u*(x,\omega'')\times$ and rearrange to

$$\begin{aligned}
&\left[-\omega''^2 + \omega^2 + i\sigma\omega\right]C(\omega,\omega'') \\
&\quad - \int dx \int d\omega' C(\omega,\omega')u*(x,\omega'')(\omega^2 + i\sigma\omega)\left[\varepsilon(x,\omega) - \varepsilon(x,\omega')\right]u(x,\omega') \\
&\quad = c^2 \int dx\, u*(x,\omega'')\tilde{J}(x,\omega)
\end{aligned} \tag{B3}$$

The integrand $C(\omega,\omega')[\varepsilon(x,\omega) - \varepsilon(x,\omega')]$ is small if dispersion is weak. This can be seen by first assuming that there is no dispersion. Then the integral is zero, and we have

$$C(\omega,\omega'') = c^2 \frac{\int dx'\, u*(x',\omega'')\tilde{J}(x',\omega)}{\left[\omega^2 - \omega''^2 + i\sigma\omega\right]} \tag{B4}$$

As $\sigma$ goes to zero, $C(\omega,\omega'')$ becomes a delta function. This localizes the integrand in (B3) to near $\omega = \omega'$, where $[\varepsilon(x,\omega) - \varepsilon(x,\omega')]$ is small, implying that when $\sigma$ goes to zero, the integral term in (B3) can indeed be dropped.

Using (B4) to represent $C(\omega,\omega'')$, (B2) can be written in the form

$$\tilde{E}(x,\omega) = \int dx' \, \tilde{K}(x,x',\omega) \tilde{J}(x',\omega) \quad, \tag{B5}$$

where

$$\tilde{K}(x,x',\omega) = c^2 \int_{-\infty}^{\infty} d\omega' \frac{u(x,\omega')u*(x',\omega')}{\left[\omega^2 - \omega'^2 + i\sigma\omega\right]} \quad. \tag{B6}$$

To find the form of the solution in the long-time, far-field limit, plug Eq.(B6) into (B5), and Fourier transform to the time domain via a contour integral, to give (after taking $\sigma$ to zero at the end)

$$E(x,t) = \int_{-\infty}^{\infty} dx' \int_{-\infty}^{\infty} dt' \, J(x',t') K(x,x',t-t') \quad, \tag{B7}$$

where the Green function is

$$K(x,x',\tau) = c^2 \int_{-\infty}^{\infty} d\omega \, u(x,\omega)u*(x',\omega)f_\omega(\tau) \quad, \tag{B8}$$

with

$$f_\omega(\tau) = -\Theta(\tau)\frac{\sin\omega\tau}{\omega} \cong \Theta(\tau)\frac{\exp(-i\omega\tau)}{2i\omega} \quad. \tag{B9}$$

The last step is a rotating-wave approximation, valid for fields whose bandwidth is much smaller than the optical carrier frequency. We consider the results in the physically relevant case that $\tau \geq 0$ and the observation point $x$ is outside the medium. Plug (B8) and (B9) into (B7), which can then be written as

$$E(x, t \to \infty) = (2\pi)^{-1} \int_{-\infty}^{\infty} d\omega \, e^{-i\omega t} \tilde{E}_+(x, \omega) \qquad , \qquad \text{(B10)}$$

where

$$\tilde{E}_+(x, \omega) = \int_{-\infty}^{\infty} dx' \, \tilde{K}_+(x, x', \omega) \tilde{J}(x', \omega) \qquad , \qquad \text{(B11)}$$

and

$$\tilde{K}_+(x, x', \omega) = \frac{-i\pi c^2}{\omega} u(x, \omega) u *(x', \omega) \qquad . \qquad \text{(B12)}$$

This is the result given in Eq.(18).

## APPENDIX C: DISTRIBUTED MICROCAVITY MODES

Write Eq.(55) as (suppressing the frequency label)

$$A'(x) = i\kappa \, B(x) e^{i\Delta x}$$
$$B'(x) = -i\kappa \, A(x) e^{-i\Delta x} \qquad \text{(C1)}$$

At the boundaries, $A'(L) = i\kappa \, B(L) e^{i\Delta L}$, $B'(0) = -i\kappa \, A(0)$. Equation (C1) is separated into two equations,

$$A''(x) = i\Delta A'(x) + \kappa^2 A(x)$$
$$B''(x) = -i\Delta B'(x) + \kappa^2 B(x) \qquad . \qquad \text{(C2)}$$

The derivative boundary conditions are, from (C1),

$$A'(L) = i\kappa B(L)e^{i\Delta L}, \quad B'(0) = -i\kappa A(0). \tag{C3}$$

The solutions are

$$A(x) = Q(x)A(0) + i\kappa P(x)B(L)e^{i\Delta L}$$
$$B(x) = -i\kappa W(x)A(0) + V(x)B(L) \tag{C4}$$

where

$$Q(x) = e^{SL+(i\Delta-S)x/2}\left[S(e^{S(x-L)}+1) - i\Delta(e^{S(x-L)}-1)\right]/D$$
$$P(x) = 2e^{(i\Delta-S)(x-L)/2}(e^{Sx}-1)/D$$
$$V(x) = e^{-(i\Delta+S)(x-L)/2}\left[S(e^{Sx}+1) + i\Delta(e^{Sx}-1)\right]/D$$
$$W(x) = 2e^{SL-(i\Delta+S)x/2}(e^{S(x-L)}-1)/D \tag{C5}$$

with

$$D = i\Delta(e^{SL}-1) + S(e^{SL}+1)$$
$$S = \sqrt{4\kappa^2 - \Delta^2} \tag{C6}$$

Energy conservation is represented by $|\kappa P(L)|^2 + |V(0)|^2 = 1$.

The cavity is modeled as a DBR and an ideal thin mirror, with field transmission and reflectivity $(\tau, \rho)$, separated by distance $d$, as in Fig.3. If a wave is incident only from the left, $B(L) = 0$, the solutions in Regions 1 and 2 are, respectively:

$$E_1(x) = A_1(x)e^{ikx} + B_1(x)e^{-ikx}$$
$$E_2(x) = A_2 e^{ikx} + B_2 e^{-ikx} \tag{C7}$$

with

$$A_1(x) = Q(x)A(0), \quad B_1(x) = -i\kappa W(x)A(0) \tag{C8}$$

This is now a standard input/output calculation for a two-mirror cavity. We can solve it by applying:

$$B_3 = -\rho A_3 + \tau r A_2 e^{ik2d}$$
$$A_2 = \tau A_3 + \rho r A_2 e^{ik2d}$$

(C9)

Solutions are:

$$A_2 = \frac{\tau}{f} A_3, \quad B_3 = R A_3,$$
$$f = 1 - \rho r e^{ik2d}, \quad R = -\rho + e^{ik2d} \frac{r\tau^2}{1 - \rho r e^{ik2d}}$$

(C10)

$B_3$ is the output field given in terms of the input field $A_3$.

The field in the DBR is found by using:

$$A(0) = A_2 e^{ikd}$$
$$B(0) = r A_2 e^{ikd}$$

,

(C11)

where $r$ is again $r = -i\kappa W(0)$. Putting all together gives in the DBR:

$$A_1(x) = Q(x) e^{ikd} \frac{\tau}{f} A_3$$
$$B_1(x) = -i\kappa W(x) e^{ikd} \frac{\tau}{f} A_3$$

(C12)

An independent solution corresponds to the case of an input, , incident only from the right. In this case the solution is

$$B(0) = \frac{t}{f}B(L), \ A(L) = R'B(L),$$

$$B_3 = \tau e^{ikd}B(0), \ R' = re^{i\Delta L} + e^{ik2d}\rho t^2 / f \qquad \text{(C13)}$$